\documentstyle[onecolumn,psfig,fleqn,amstex,rotating]{mn}
\newcommand{\hi}{\mbox{H\ {\footnotesize I}}}

\newcommand{\nhi}{\mbox{\scriptsize H\ {\tiny I}}}

\def\etal{{et al.\ }}

\def\lsim{~\rlap{\raise 0.4ex\hbox{$<$}}{\lower 0.7ex\hbox{$\sim$}}~}
\def\gsim{~\rlap{\raise 0.4ex\hbox{$>$}}{\lower 0.7ex\hbox{$\sim$}}~}

\def\tcbr{T_{_{\rm CBR}}}
\def\ts{T_{\rm s}}
\def\tk{T_{\rm \; K}}
\def\tb{T_{\rm b}}
\def\xhi{x_{_{\rm HI}}}

\def\l0{L_\ast(0)}

\def\s0{S_\ast(0)}

\def\omg0{\Omega_0}

\def\a2{\alpha^{(2)}}
\def\ahi{\alpha_{\nhi}^{(2)}}

\def\mpc3{\ {\rm Mpc^{-3}}}
\def\gpc3{\ {\rm Gpc^{-3}}}

\def\hmpc{\ {\rm h^{-1}Mpc}}

\begin{document}

\title[]
{The spin temperature of  neutral hydrogen during cosmic pre-reionization}
\author[Nusser]{Adi Nusser$^{1,2}$\\\\
$^{1}$Physics Department and the Asher Space Researrch Institute, Technion, Haifa 32000, Israel\\ $^2$The Institute of Astronomy, Madingley Rd, Cambridge CB3 OHA\\}
\maketitle

\begin{abstract}
We re-examine the role of collisions in decoupling the \hi\ 21-cm spin temperature
from the cosmic microwave background (CMB).
The cross section for de-exciting the 21-cm trasition in  collisions 
with  free electrons is more than 10 times 
larger than it is in  collisions with other atoms. 
If the fraction of  free  electrons in the diffuse cosmic gas
is between 10 and 30 per cent then collisions 
alone can decouple the spin temperature from the cosmic microwave background (CMB), 
even in moderately under-dense regions at  $z\gsim 15$.  
This  decoupling is especially important during the very early 
stages of re-ionization when a Ly$\alpha$ continuum background 
had yet to be established. 
As a detailed  example, we develop a semi-analytic model to quantify 21-cm emission signatures 
from a diffuse gas which is  partially ionized at $z\gsim 10$ by an
X-ray background.
We find 21-cm  differential brightness temperature fluctuations with a mean of $\gsim 8\rm \; mK$ and 
 a $rms$ value as large as $5\rm \; mK$, for a frequency resolution bandwidth 
of $100\rm \; KHz$ and a beamsize of 3 arcmin.
Another example where free electron-atom collisions are important 
is during the recombination of bubbles ionized by short-lived UV sources. When
the ionized fraction in these bubbles  drops 
to  10-20 per cent their 
differential temperature 
can be as high as $10\rm \; mK $. 

\end{abstract}



\section {Introduction}
\label{sec:introduction}

Recent observational data of the universe have substantially tightened our grip 
on the cosmological model and its fundamental parameters. 
 Most prominent of these data are 
maps of the cosmic microwave background (CMB), in particular those measured by  the {\it Wilkinson Microwave Anisotropy Probe} (WMAP) satellite (e.g. Spergel \etal 2003), the 2dF and SDSS galaxy redshift surveys
(Percival \etal 2002; Zehavi \etal 2002), and the Ly$\alpha$ forest seen in spectra of
quasars (QSOs) (Croft \etal 1999, 2002; McDonald \etal 2000 \& 2004, Nusser
\& Haehnelt 1999, 2000; Kim \etal 2004; Viel,  Weller \& Haehnelt 2004).
Despite the remarkable recent achievements, our understanding of the 
initial stages of the formation of structure remains poor. 
The weakest link  is the era  
of hydrogen 
re-ionization at high redshifts. 
The WMAP polarization measurement (e.g. Spergel \etal 2003) imply a universe  that 
must have been ionized at redshifts $z>13$.  However, 
the imprinted signal of re-ionization in the CMB is insensitive to the 
details of the initial stages. A  probe of these details can be found in 
\hi\ absorption features in spectra of high redshift QSO.
These features can reveal the spatial distribution of ionized  
 regions as a function of redshift (Becker \etal 2001, Fan \etal 2002).
Unfortunately, bright QSOs are hard to detect at high redshifts and, also,  the inference of constraints on the three-dimensional structure of re-ionization can 
be very tricky (e.g. Nusser \etal 2002). 
A more promising probe of reionization, especially of the initial stages, 
is observations of the redshifted 21-cm emission/absorption lines 
produced by  \hi\ in the high redshift universe (Field 1959, Sunyaev \& Zel'dovich 1975, Hogan \& Rees 1979, Subramanian \&  Padmanabhan 1993). 
The 21-cm line is produced in the transition between the triplet and
singlet sublevels of the  hyperfinestructure of the ground level of neutral hydrogen atoms. 
This wavelength  corresponds to a frequency of 1420MHZ
and a temperature of $T_{*}=0.068{\rm \; K}$.  
If $n_{1}$ and $n_{0}$ are the populations of the triplet and 
singlet ground state of HI atoms then the spin temperature is defined by $n_{1}/n_{0}=3 \exp(-T_{*}/\ts)$.
The  CMB radiation tends to bring $\ts$ to its own temperature of $\tcbr\approx 2.73(1+z)\rm \; K$ (Mather \etal 1994). An \hi\ region would
be visible against the CMB either in  absorption if $\ts<\tcbr$ or emission 
 if $\ts>\tcbr$. 
Two mechanisms for decoupling the spin and CMB
temperature have been proposed (Field 1958).
The first is the Wouthuysen-Field process (Wouthuysen 1952, Field 1958) which is also termed ``Ly$\alpha$ pumping''. In this process the hyperfine 
sublevels  are interchanged  by
the absorption and emission of a Ly$\alpha$ photon. The final
product of this process is a spin temperature that equals the ``color''
temperature of the Ly$\alpha$ photons.
The color temperature eventually approaches the 
thermal temperature of the gas as a result of recoil effects by repeated scattering off the \hi\ atoms (Field 1959). 
The other mechanism is spin exchange by means collisions of atoms with 
free electrons and other atoms.  Both of these effects work towards  bringing the spin 
temperature close to the gas thermal temperature\footnote{There is actually another  decoupling mechanism. Immediately after hydrogen atoms recombine they populate the triplet and singlet hyperfine sublevels according to the ratio $3:1$, leading to $\ts=\infty$.
However, this decoupling is just a transient  as very rapidly 
the  CMB radiation field  
brings $\ts$ to finite values.}.
The efficiency of Ly$\alpha$ pumping  has been studied thoroughly by 
several authors (Scott \& Rees 1990, Madau, Meiksen \& Rees 1997, Ciardi \& Madau 2003, Tozzi \etal  2000, Chen \& Miralda-Escud\'e 2004). 
For reionization by stars alone the  Ly$\alpha$ pumping is believed to be  very efficient
even when the volume filling factor of ionized regions is 
less than a few per cents.
The reasoning behind that is as follows.
During the early stages of reioinization all UV photons  emitted with $h\nu>13.6\; \rm ev$ 
by a  stellar source are consumed in ionizing the surrounding gas (mainly  \hi ). 
A  Ly$\alpha$ continuum photon  with $10.2<h\nu<13.6\; \rm ev$
is free to travel until it is redshifted to a $h\nu \approx 10.2\; \rm ev$  where it 
is captured in the Ly$\alpha$ line of  a ground state  \hi\ atom.   Such a photon travels
a comoving distance of  $R_{\alpha}\sim 6000\hmpc (1+z)^{-1/2}(1-(4(h\nu/13.6)/3)^{1/2})$ (about 350$\hmpc$ at $z=20$ for $h\nu=13.6$ ev) before
its capture. Since  a star emits at least 4 times more photons in the Ly$\alpha$ continuum than in the UV  (e.g. Ciardi \& Madau 2003)),  one expects a significant amount of 
the  continuum 
photons to reach the \hi\ far beyond  
ionized regions. According to Ciardi \& Madau (2003)  Ly$\alpha$ pumping should already 
have become efficient when the volume filling factor of ionized regions 
is as small as a few per cent.  
This conclusion is invalid when the ionizing sources
are highly clustered. The Ly$\alpha$ continuum from  an aggregation of sources 
 could decouple the spin-CMB temperature 
in a sphere of maximum radius of  $R_{\alpha}$ around it, independent of  the number 
of  Ly$\alpha$ continuum photons emitted  by these sources. However, the ionized region 
around this aggregation is proportional to the number of 
sources in it. 
Further,  X-ray photons from accreting black holes could easily 
have played an important role in an early re-ionization epoch (e.g. Ricotti \& Ostriker 2004).
Because of the large uncertainties in describing reionization (c.f. Cen 2003, 
Wyithe \& Loeb 2003) it is prudent to examine 
in detail the consequences of other mechanisms for decoupling the spin-CMB
temperatures.

In the absence of free electrons,   spin  exchange by \hi\ atom-atom collisions 
is important only in regions with high density contrasts,
$\delta >20[(1+z)/10]^{-2}$ (e.g. Ciardi \& Madau 2003). 
This condition is satisfied in minihalos with  
$\tcbr\ll T_{\rm vir}\lsim 10^{4}{\rm \; K}$ 
which could contain a significant 
fraction of \hi\ at $z\sim 8$ (Iliev \etal 2003) in the absence of 
external heating sources.
 In this paper we argue that collisions of \hi\ atoms with free electrons
  and other atoms in 
moderate density regions with 70-90 per cent neutral fraction can induce  significant 21-cm emission against the CMB.
As an  example of this we focus on 
  pre-reionization (i.e. early stages of reionization) by X-ray photons  (Oh 2001, Venkatesan, Giroux \& Shull 2001, Machacek, Bryan \& Abel 2003, Ostriker \& Gnedin 1996, Ricotti \& Ostriker 2004, Ricotti, Ostriker \& Gnedin 2004). X-ray photons have 
such a small  cross section for \hi\ ionization  so that they 
rapidly form a uniform ionizing background (e.g. Ricotti \& Ostriker 2004).
The average ionized fraction in the presence of this background is 
in the range 5-20 per cent for $30\gsim z\gsim 10$, which is consistent with the 
current constraints on the observed soft X-ray background (Dijkstra \etal 2004).
Under these circumstances, collisions are very efficient 
at decoupling the spin-CMB temparetures. 
We  develop 
a simple  semi-analytic model to estimates the 21-cm emission fluctuations
induced by  spin-CMB decoupling by means of collisions in the framework of the X-ray scenario.
We also consider  21-cm emission during the recombination of 
``bubbles'' ionized by sporadic short-lived sources
of UV ionizing radiation.  
Gas in a newly formed ionized bubble would have   
a temperature of a few $10^{4}\rm \; K$ and 
would immediately  start cooling (mainly by Compton cooling) and recombining
into \hi .
Electron-atom collisions inside the bubble maintain the recombined \hi\ at 
$\ts\gg\tcbr$.
This high $\ts$ amounts to a significant 21-cm emission against the CMB when  the \hi\ fraction  is  the range $70-90$ per cent. This emission could last for  $\sim 10^{8}\rm Yr$ so that repetitive formation of these bubbles could yield an appreciable signal.

The outline of the paper is as follows. In \S 2 we summarize the basic equations.
In \S 3 we present results for the 21-cm differential brightness temperature 
at mean gas density as a function of temperature, the \hi\ fraction, and redshift. 
The temperature fluctuations are  estimated numerically in \S 4. With the exception 
of subsection 4.3, all of \S 4 is devoted to the role of collisions 
in the X-ray  pre-reionization   
scenario. We discuss the results and conclude in \S 5.

\section{The basics}
We assume throughout that  spin exchange by means of collisions is the dominant mechanism for 
decoupling the spin temperature from 
the CMB. In this case 
The spin temperature is given by (Field 1959),
\begin{equation}
\ts=\frac{\tcbr+y_{\rm c}T_{\rm \; K}}{1+y_{\rm c}} \; ,
\label{tspin}
\end{equation}
where
\begin{equation}
y_{\rm c}=\frac{C_{10}}{A_{10}} \frac{T_{*}}{T_{\rm \; K}} \; ,
\label{yc}
\end{equation}
with $C_{10}$ denoting the total de-excitation rate as a result of 
collisions with free electrons and other atoms. Following Field (1959), we write
\begin{equation}
y_{\rm c}={\tilde y}_{_{\rm HI}}n_{_{\rm HI}}+ {\tilde y}_{_{\rm e}}n_{_{\rm e}}
\; .
\end{equation}
In the following we adopt  the values of ${\tilde y}_{_{\rm HI}}$ and ${\tilde y}_{_{\rm e}}$ listed in Table 2 of Field (1958). The corresponding values for $C_{10}$ are 
in good agreement with other published data for the relevant temperature
range considered here (Allison \& Dalgarno 1969). At $\tk=3000\rm \; K$ and 
$10^{4}\rm \; K$ the ratios of ${\tilde y}_{_{\rm e}}$ to ${\tilde y}_{_{\rm HI}}$ are
 20 and 13.8, respectively.

Intensities, $I(\nu)$  at radio frequency are expressed in terms of brightness temperature
defined as $\tb=I(\nu)c^{2}/2k\nu^{2}$, where $c $ is the speed of light and $k$ is Boltzman's constant (Wild 1952).
The {\it differential brightness temperature} (hereafter, DBT) against the CMB
of a small patch of gas with $\ts$ at redshift $z$ is (e.g. Ciardi \& Madau 2003),
\begin{equation}
\delta \tb=16{\rm \; mK} \; x_{_{\rm HI}}(1+\delta)\left(1-\frac{\tcbr}{\ts}\right)
\left(\frac{\Omega_{\rm b} h}{0.02}\right) 
\left[  \left( \frac{1+z}{10} \right) \left(\frac{0.3}{\Omega_{\rm m}}\right)
\right]^{{1/2}}\; . 
\label{tb}
\end{equation}
where   $\delta$ and  $x_{_{\rm HI}}$ are the gas density contrast and 
the fraction of \hi\ in the patch, respectively.

\section{The  brightness temperature at mean density}

\begin{figure}
\centering
\mbox{\psfig{figure=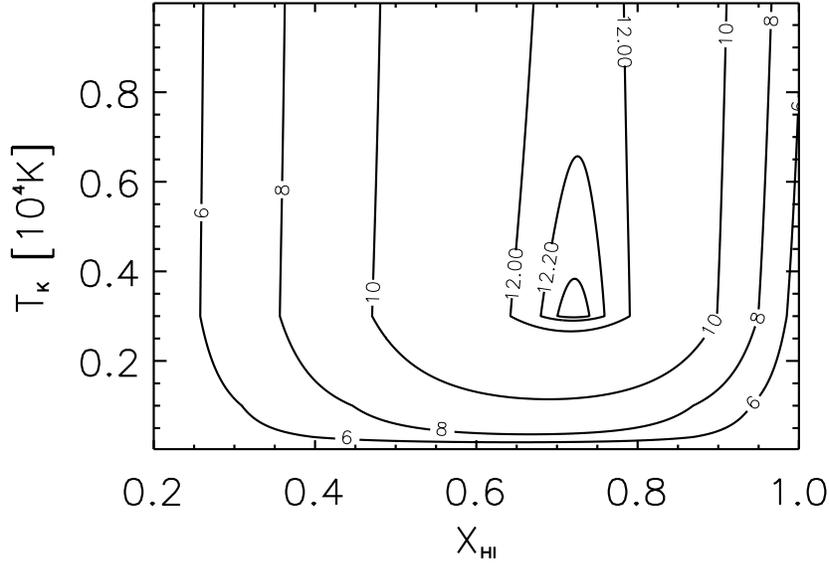,height=3.5in,width=5in,bbllx=54bp,bblly=360bp,bburx=558bp,bbury=720bp,clip=}}
\caption{Contour map of the DBT as a function of the neutral fraction, $\xhi$, and  the gas thermal temperature, $T_{\rm \; K}$. The calculation is for a gas at mean density at redshift $z=15$. The DBT levels (in mK) are indicated by the contour labels.
} 
\label{fig:one0}
\end{figure}

We first study the DBT, $\delta \tb$, at mean density ($\delta=0$) as a function of redshift.
We assume throughout that the spin and CMB temperatures are decoupled only by means of 
electron-atom and atom-atom collisions.
In Fig.~\ref{fig:one0} we show a contour map of 
$\delta \tb$ at mean gas density ($\delta=0$)  at $z=15$ as a function of $\xhi$  and the gas temperature $T_{\rm \; K}$.  The DBT peaks  at $\xhi\approx 0.72$ and
$T_{\rm \; K}\approx 3000\rm \; K$. It  is insensitive to  $T_{\rm \; K}$ 
above 3000K, but varies appreciably as a function of $\xhi$. For
 $ 10^{4}\rm \; K \lsim T_{\rm \; K}\lsim 3\times 10^{4}\rm \; K$ there is an increase by  a nearly factor of $2$ between $\xhi=1$ and 0.7.

\begin{figure}
\centering
\mbox{\psfig{figure=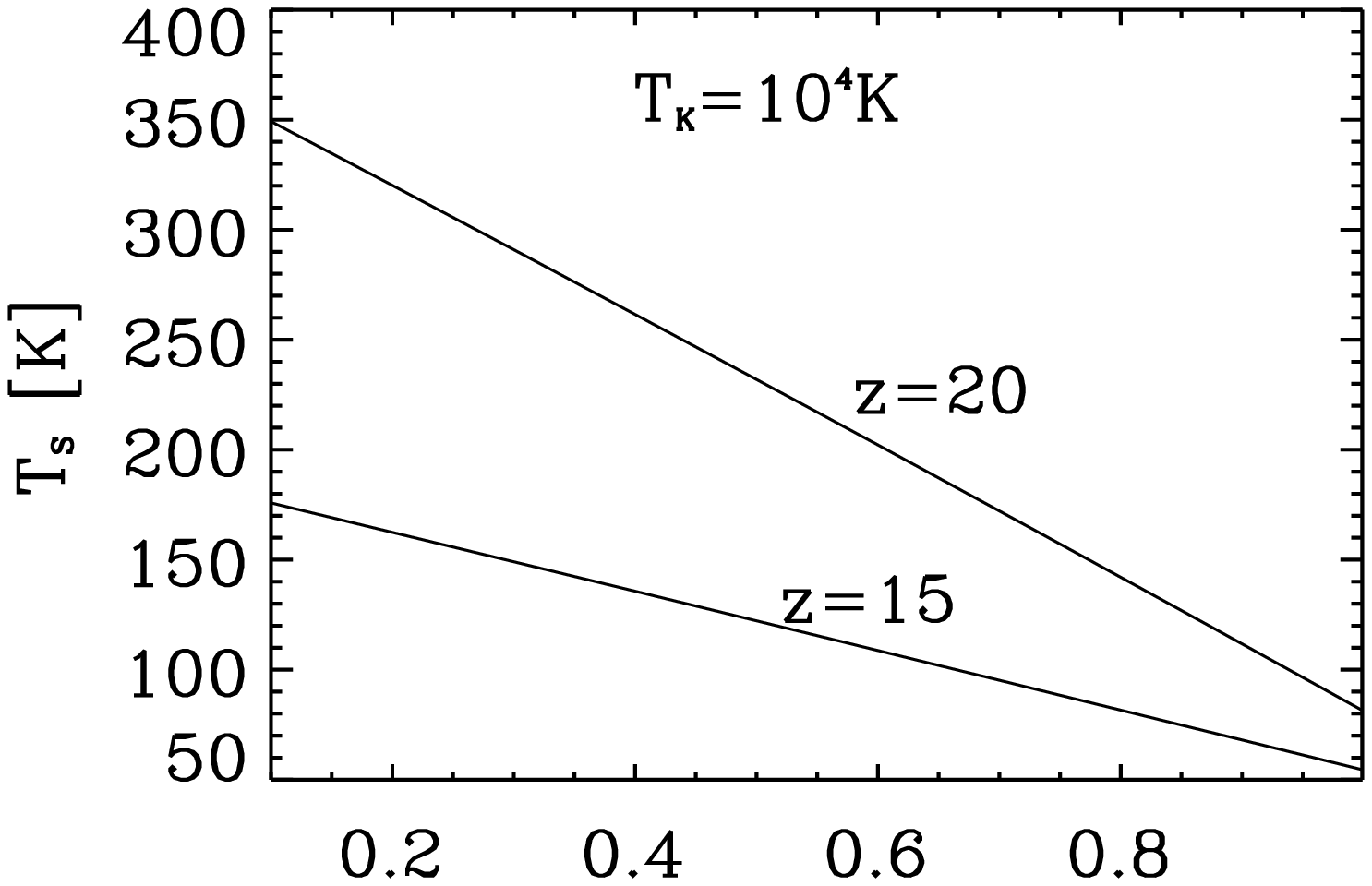,height=3.in,width=4.5in,clip=}}
\mbox{\psfig{figure=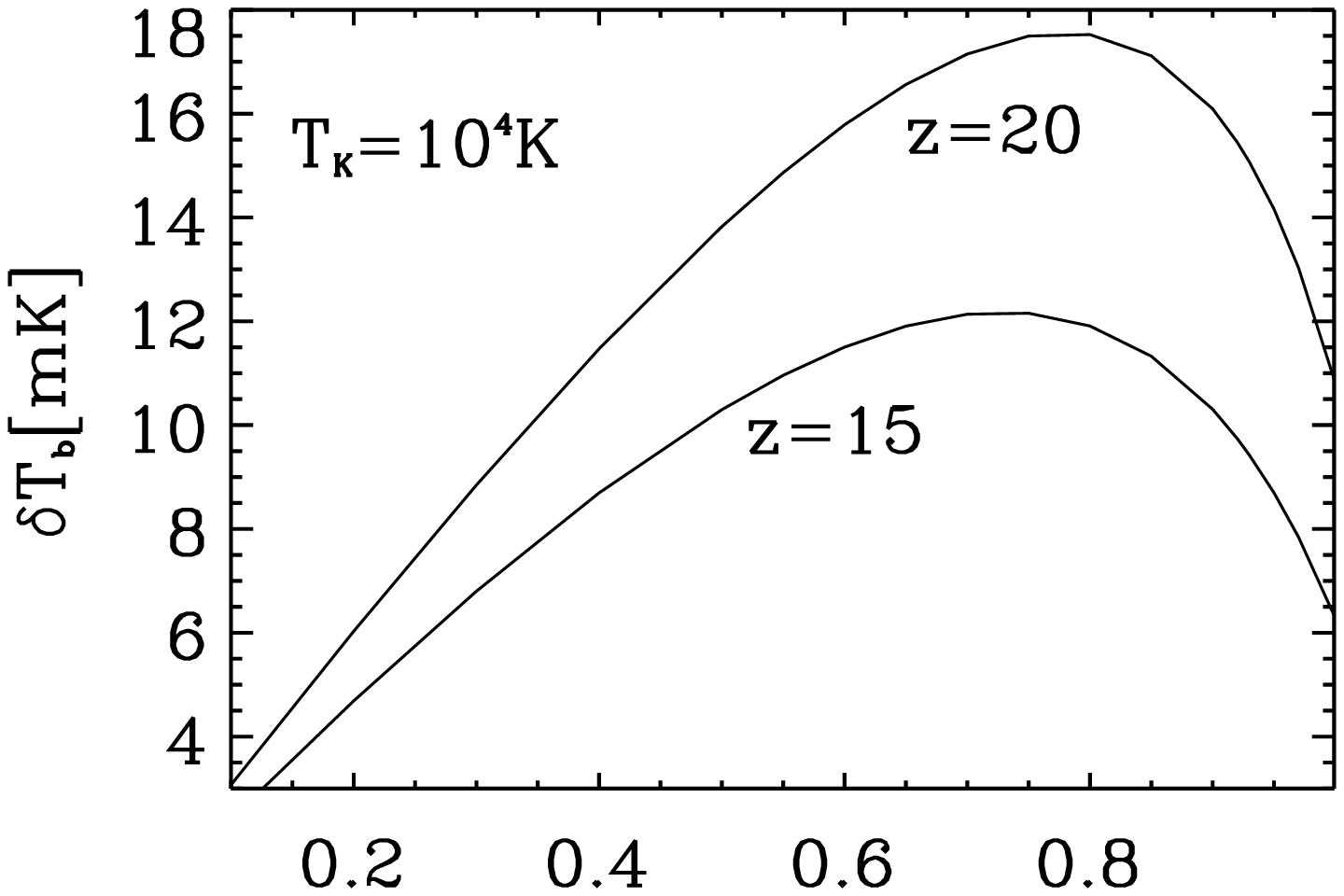,height=3.in,width=4.5in,clip=}}
\mbox{\psfig{figure=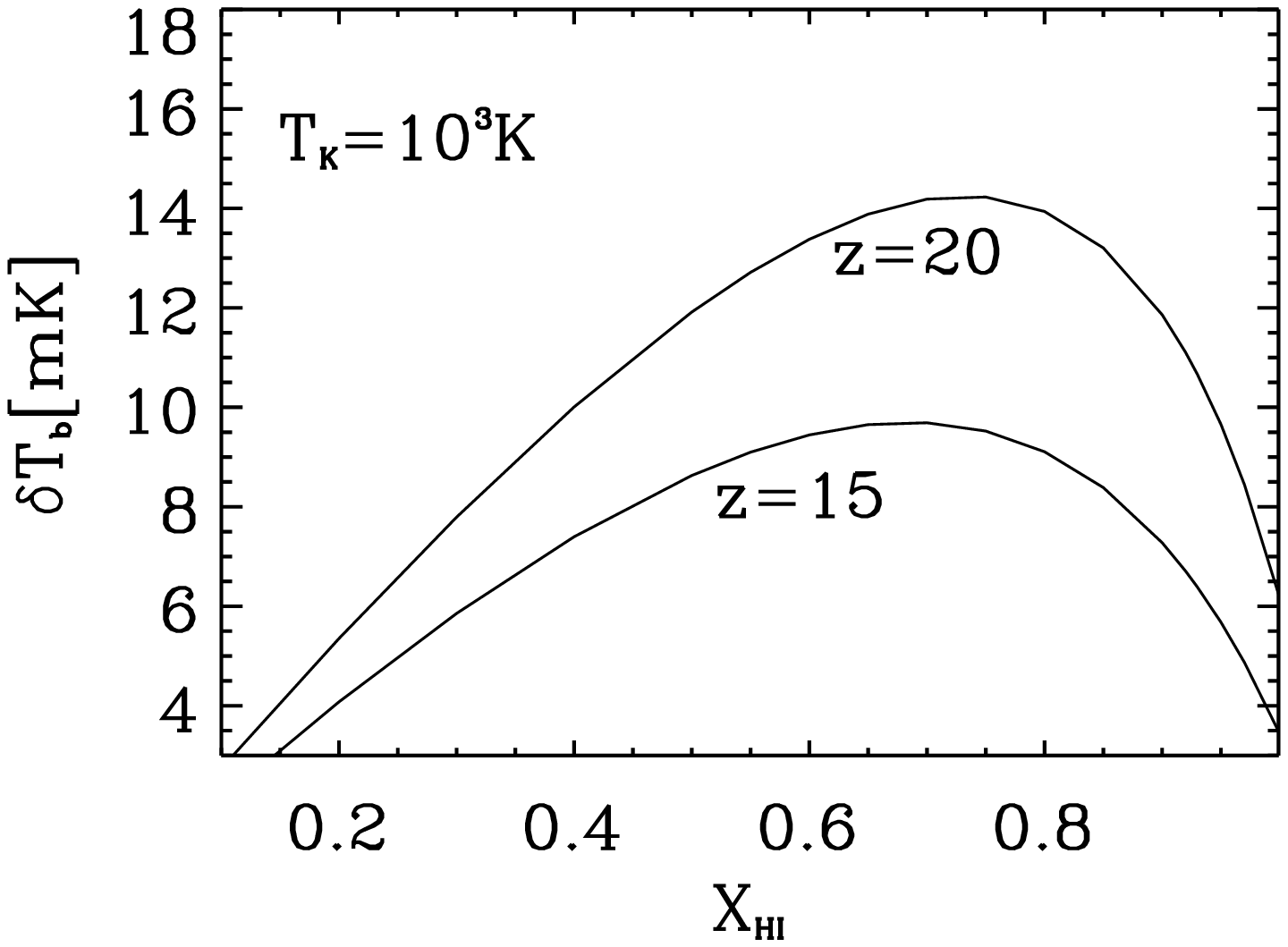,height=3.in,width=4.5in,clip=}}
\caption{ {\it Top}: The spin temperature at mean density as a function of the neutral fraction  $\xhi$ for $T_{\rm \; K}=10^{4}\rm \; K$ at redshifts $z=15$ and $20$ as indicated in the figure.
{\it Middle}: The DBT at mean density  as a function of $\xhi$ 
for  $T_{\rm \; K}=10^{4}\rm \; K$. {\it Bottom}: The DBT for $10^{3}\rm \; K$.} 
\label{fig:one}
\end{figure}

\begin{figure}
\centering
\mbox{\psfig{figure=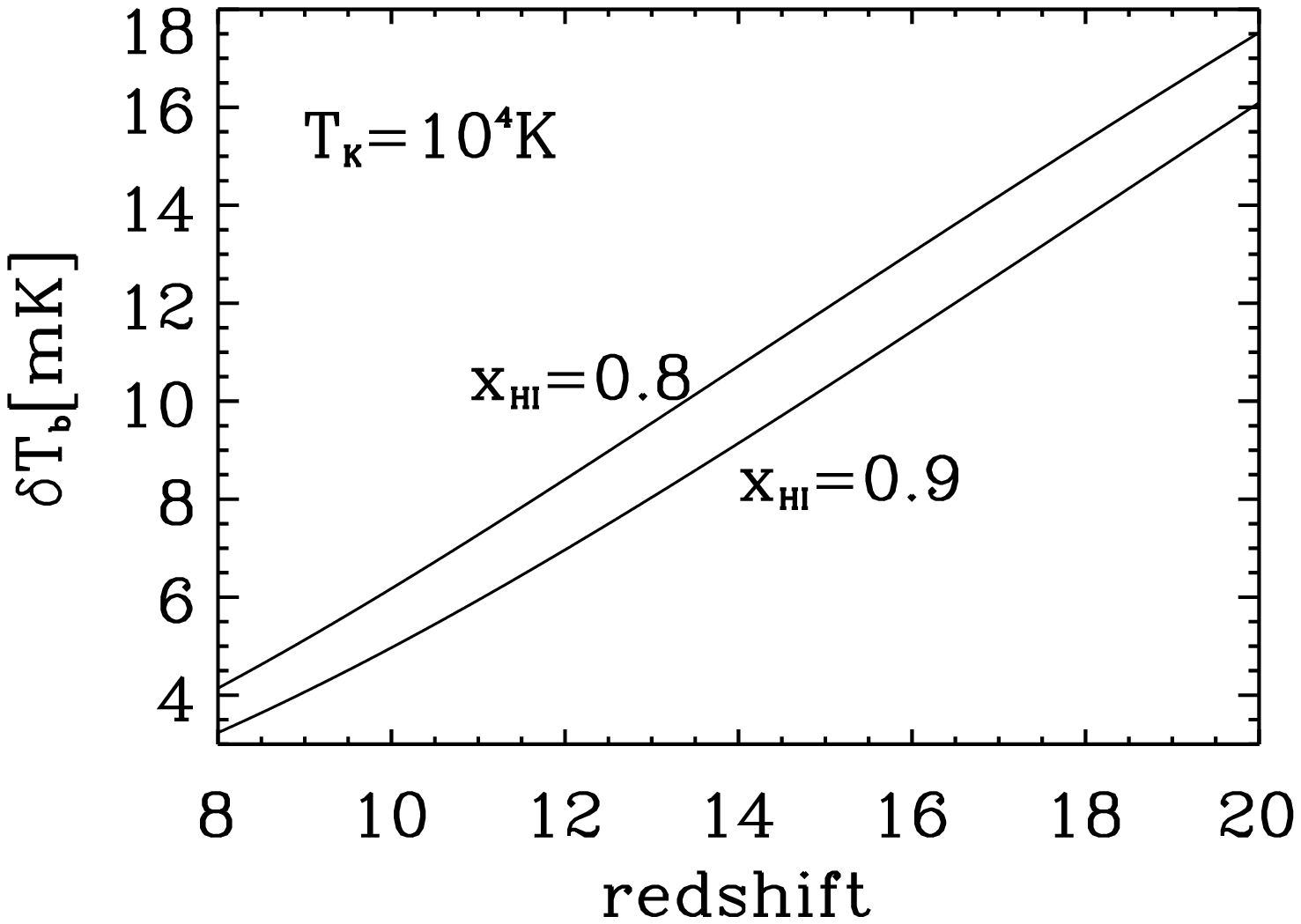,height=3.5in,width=5in}}
\caption{The DBT at mean density as a function of redshift, plotted for
two values of $\xhi$ as indicated in the figure. The gas temperature is
 $10^{4}\rm \; K$.}
\label{fig:two}
\end{figure}

In the top panel of Fig.~\ref{fig:one} we examine the spin temperature, $\ts$, as a function of the neutral fraction, $\xhi$. This temperature  is  
very sensitive to $\xhi$ because of  the enhanced coupling between 
$\ts$ and $\tk$ produced by electron-atom collisions.
It is well below  $\tk$, but  
significantly larger than the CMB temperature $\tcbr=2.73(1+z)$.
The middle and bottom panels in Fig.~\ref{fig:one} show $\delta \tb$ as 
a function of $\xhi$ for $\tk=10^{4}\rm \; K$ and $10^{3}\rm \; K$, respectively.
Here also we see that a small free electron fraction can substantially boost
up $\delta \tb$. 
For $\xhi\approx 1$  collisions with free electrons become less
effective at decoupling $\ts$ from $\tcbr$, making  $\delta \tb\propto 
(1-\tcbr/\ts)$  (see Eq.~\ref{tb}) drop by a factor of 2 from its peak value at $\xhi=0.8$.
Note that the  $\delta \tb$ curves peak at $\xhi=0.7-0.9$, close to the 
values expected in the X-ray pre-reionization scenario  (Ricotti \& Ostriker 2004).
In Fig.~\ref{fig:two} we show the redshift evolution of $\delta \tb$ at
$\delta=0$ for $\xhi=0.8$ and 0.9, as indicated in the figure. 
In a realistic scenario for X-ray pre-reionization $\xhi$  varies with
redshift (Ricotti \& Ostriker 2004). The history of X-ray pre-ionization, if indeed happened, 
could in principle be  probed by observations of the mean temperature as a function of 
observed frequency, $\nu=1420{\rm MHz}/(1+z)$. In practice, this signal maybe 
overwhelmed  by foreground 
source (Shaver \etal 1999; Baltz, Gnedin \&
Silk 1998, Ricotti, Ostriker \& Gnedin 2004).

\section{The 21-cm temperature fluctuations}

\subsection{The model}

As a concrete example for assessing the role of collisions 
we consider the DBT in the X-ray pre-reionization scenario (Oh 2001, Venkatesan, Giroux \& Shull 2001, Machacek, Bryan \& Abel 2003, Ostriker \& Gnedin 1996, Ricotti \& Ostriker 2004, Ricotti, Ostriker \& Gnedin 2004). 
The mean free path of X-ray photons is much larger than the 
correlation length of the sources and their mean separation. 
Therefore, in this scenario  X-ray photons act as a uniform ionizing background
and the \hi\ fraction, $\xhi$, at a point with density contrast, $\delta$, 
is  determined by the equation for 
 photoionization equilibrium, 
\begin{equation}
\Gamma \xhi = {\ahi} {\bar n}  (1+\delta) (1-\xhi )^{2} \; , 
\label{photo}
\end{equation} 
where $\Gamma$ is the X-ray photoionization rate per hydrogen atom, $\ahi$ is the 
recombination cross section to the second excited atomic level, and $\bar n$ is 
the mean density of total (ionized plus neutral) hydrogen.
According to Ricotti \& Ostriker (2004),  the mean \hi\ fraction should be maintained
at a level $\xhi\sim 0.8-0.9$ starting from redshifts $20-40$ in order to match 
the WMAP value of $\tau_{\rm e}\sim 0.17$ for  optical depth to Thomson scattering (Spergel \etal 2003).
Therefore, here we fix the ration $\ahi {\bar n }/\Gamma$ by 
assuming that $\xhi$ is given at mean density, i.e. at $\delta=0.$
Then $\xhi$ for any $\delta$ can be easily found  
using (\ref{photo}).

\begin{figure}
\centering
\mbox{\psfig{figure=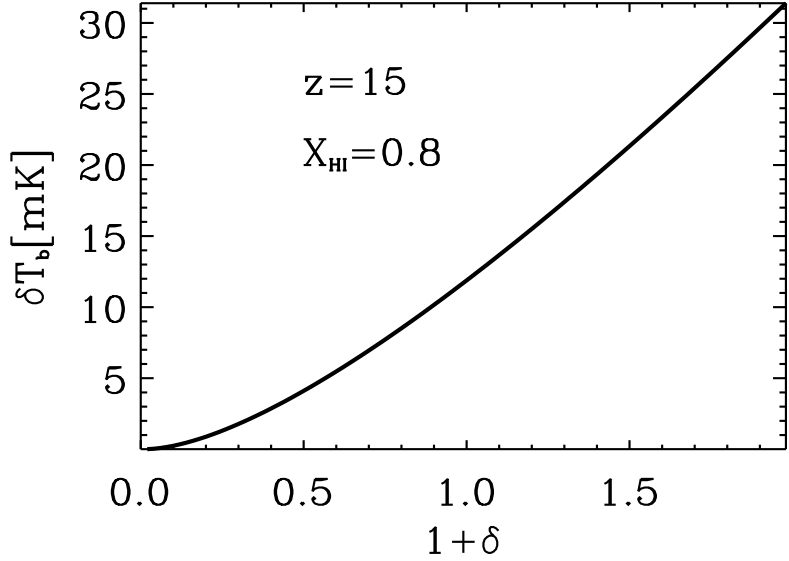,height=3.5in,width=5in}}
\caption{The DBT as a function of the gas density contrast at $z=15$. The 
 \hi\ fraction at $\delta=0$ is  $\xhi=0.8$. For $\delta \neq 0$ we compute 
 $\xhi$ assuming a uniform ionizing background as explained in the text. }
\label{fig:tb_delta}
\end{figure}

The DBT fluctuations over the sky and as a function 
of frequency contain significant cosmological information. In the case of the X-ray scenario, these fluctuations are a direct probe of the density fluctuations 
during the pre-reionization period since, as can be seen from  
Fig.~\ref{fig:tb_delta},
the DBT is an increasing function of the local density.
In the following we make an attempt at assessing  the statistical 
properties of these  fluctuations.
Ideally one would like to have a suite of hydro simulations combined with 
radiative transfer in order to compute the ionized fraction as a function 
of redshift and position for a given cosmological model (e.g. Gnedin \& Shaver  2004,
Ricotti, Ostriker \& Gnedin 2004).
Unfortunately, current simulations of this type are  limited to small boxes ($\sim 10-20\hmpc$) making them unsuitable for detailed statistical analyses. Here we adopt the
alternative approach of 
modeling the dark and baryonic matter using semi-analytic approximations. 
We work in the framework of the  $\Lambda\rm CDM$ cosmological model with 
$\Omega_{\rm m}=0.3$ and $\Omega_{\Lambda}=0.7$.
In order to generate a prediction for the 21-cm emission maps in this model
we follow the following steps,
\begin{enumerate}
\item generate a random gaussian realization in a cubic grid box. This field 
is to represent the dark matter (DM) 
density field in a $\Lambda\rm CDM$ cosmology. 
\item obtain the gas density field by filtering  the DM density with a smoothing 
window designed to mimic pressure effects in the linear regime.
\item assume that the gravitationally evolved (non-linear) gas density field is related to the linear field
by a log-normal mapping so that the probability distribution function of the 
evolved field is log-normal (Bi, Boerner \& Chu 1991).
\item compute the ionized fraction at each point in the box for a uniform X-ray background. The amplitude of the ionizing background is fixed by 
the  value of the ionized fraction at mean density.
\item compute the spin temperature at each point assuming a 
uniform gas temperature
of $10^{4}\rm \; K$ (e.g. Ricotti \& Ostriker 2004) throughout the box.
\end{enumerate}

We work with a cubic  grid of $256^{3}$ points in 
a box of size $L=40 \hmpc$ (comoving) on the side. 
In k-space the smoothing window we use to mimic pressure effects is of the 
form (Bi, Boerner \& Chu 1991),
\begin{equation}
W(k)=\frac{1}{1+\left(k/k_{\rm J}\right)^{2} }\; .
\label{jfilt}
\end{equation}
In this expression  $k_{J}$ is the Jeans scale which depends on the gas temperature and redshift as,
\begin{equation}
k_{\rm J}(z)=\sqrt{\frac{3}{2}} \frac{H(z)}{(1+z) c_{\rm s}} \; ,
\label{kj}
\end{equation}
where $c_{\rm s}$ is the speed of sound. 
We smooth the DM density using the filter (\ref{jfilt}) with $k_{\rm J}$ corresponding to 
$T_{\rm \; K}=10^{4}\rm \; K$ in order to derive the  {\it linear} gas density contrast, 
$\delta_{\rm L}$.
The log-normal model then gives the {\it non-linear} gas density, $\delta$, as (Coles \& Jones 1991, Bi 1993), 
\begin{equation}
1+\delta=\exp\left(\delta_{\rm L}-\sigma_{\rm L}^{2}/2\right) \; ,
\end{equation}
where $\sigma_{\rm L}$ is the $rms$ value of $\delta_{\rm L}$. This transformation
yields a log-normal probability distribution function for $\delta$.
The value of $\sigma_{\rm L}$ determines the amplitude of the evolved gas density. 
Here we tune $\sigma_{\rm L }$ so that the desired $rms$ value, $\sigma_{\delta}$, of $\delta$ is attained. 
There is a large uncertainty in the determination of 
$\sigma_{\delta}$ and, also, in the form of the filtering window used to infer the gas density from the DM distribution (e.g.  Bi \& Davidsen 1997, Hui \& Gnedin 1997, Nusser 2000, Viel \etal 2002).   In any case, one of our  goals is to show 
 that a detection
of the 21-cm DBT fluctuations could actually constrain the properties of the gas density field as long as $\sigma_{\delta}$ is high enough.  
We will present results for  $\sigma_{\delta}=1$ and $\sigma_{\delta}=0.6$.
These values are  inferred  from
the non-linear power spectrum of the $\Lambda \rm CDM $ computed
from the linear power spectrum using the recipe outlined in Peacock (1999).
The linear power spectrum is normalized such that the linear rms value of of density fluctuation in spheres of radius 8$\hmpc$ is $\sigma_{8}=1$.  
The non-linear power spectrum is smoothed with the filter given in (\ref{jfilt}) to derive an estimate 
for $\sigma_{\delta}$. The outcome of this is $\sigma_{\delta}$ close to 0.7.
However, because of the uncertainties in the shape of the 
filter and the value of $\sigma_{8}$ this value of $\sigma_{\delta}$ can  only be a a crude estimate. Hence we present results for 
 $\sigma_{\delta}=1$ and 0.6. 
Once the evolved density field is obtained on the cubic grid, we solve the equations
(\ref{tspin}), (\ref{tb}), and (\ref{photo}) to derive the DBT, $\delta \tb$,
at each grid point.

\subsection{The fluctuations}

We begin with analyzing two dimensional maps of the DBT.  
The beamsize, $\Delta \theta$, and the frequency bandwidth, $\Delta \nu$, determine
the resolution at which ``an observed map'' represents the underlying temperature
fluctuations.  We ignore 
foreground contamination  and refer the reader to Oh \& Mack (2003) and
Di Matteo \etal (2002) for details.  We also do not include the effect of redshift distortions in the analysis (e.g. Nusser 2004).
 Here we mimic observations of DBT  as follows. We place the box at the desired redshift and arbitrarily pick the plane $x_{1}-x_{2}$ in the box to represent the
plane of the sky in a given direction. Using the angular diameter distance, 
we express the coordinates $x_{1}$ and $x_{2}$  
 as angular positions in the sky.  
The  $x_{3}$ coordinate  is expressed in terms of  the frequency $1420{\rm MHz}/[1+z(x_{3})]$, where $z(x_{3})$ is the redshift corresponding to $x_{3}$.
The DBT in the box is then given as a function of the 
angular position (in the $x_{1}-x_{2}$ plane) and frequency (the $x_{3}$ direction).  
At redshift 15 our  box  ($L=40\hmpc$)
represents a region spanning a 2.59MHz range in frequency and subtending $19.22$ arcmin on the sky.  
We  smooth the  DBT field with an  ellipsoidal  gaussian window
having of a FWHM of $\Delta \nu $ in the frequency axis,
and a FWHM of $\Delta \theta$ in the perpendicular (angular) plane.

\begin{figure}
\centering
\mbox{\psfig{figure=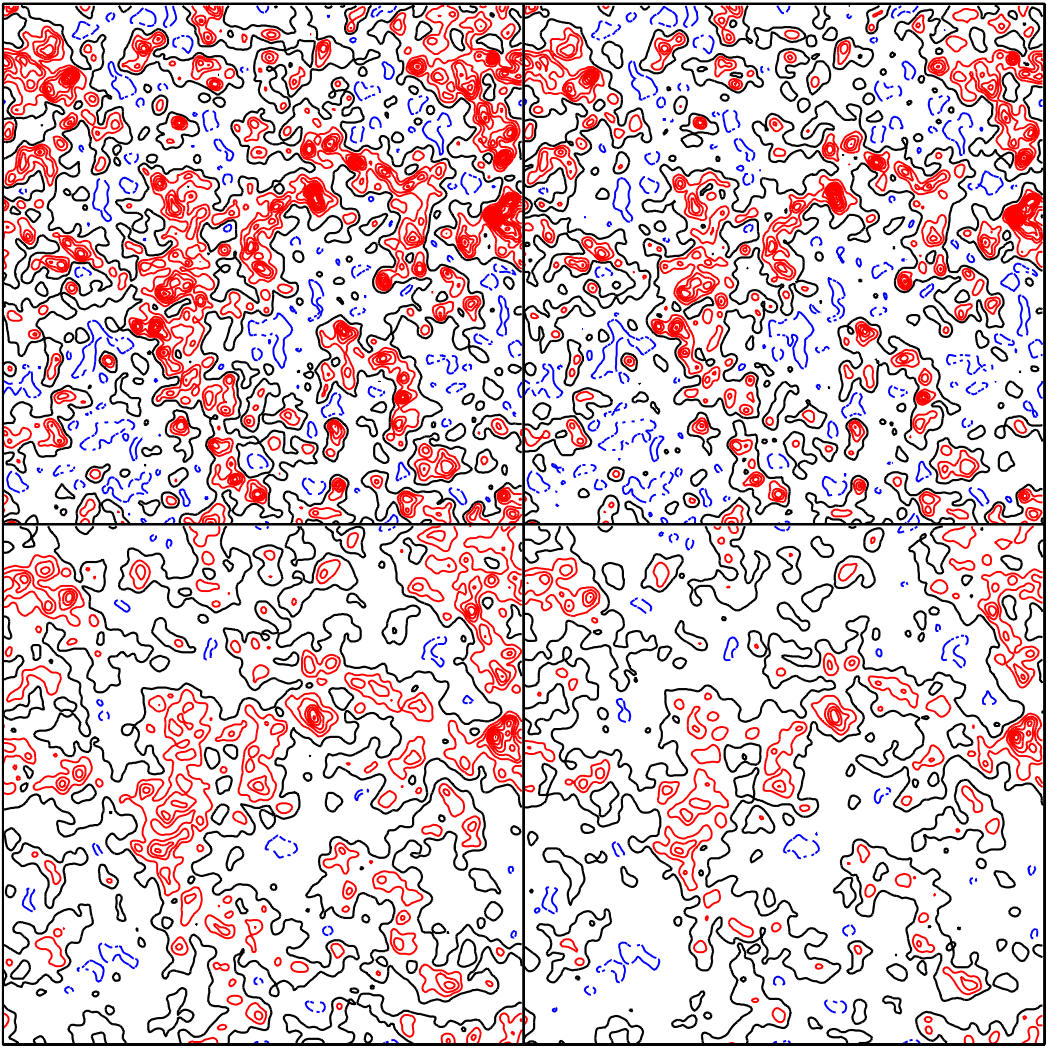,height=6.5in,width=6.5in}}
\caption{Contour map of the fluctuations in the DBT estimated from 
the gas density field with $\sigma_{\delta}=1$ at z=15. The panels in the left and right columnn, respectively, correspond to  $\xhi=0.8$ and $0.9$ at $\delta=0$. The top and bottom panels are for frequency bandwidth of $\Delta \nu=10\rm \; KHz$ and $100\rm \; KHz$, respectively. 
The maps are for an angular resolution of $3.4$ arcmin and the box size is $40\hmpc$ on the side. The solid thick line marks the 8K level of the DBT. Fluctuations below and above the 8K level are represented by the light solid and dotted lines, respectively. The contour spacing is $5\rm \; mK$. }
\label{fig:tmap1}
\end{figure}

\begin{figure}
\centering
\mbox{\psfig{figure=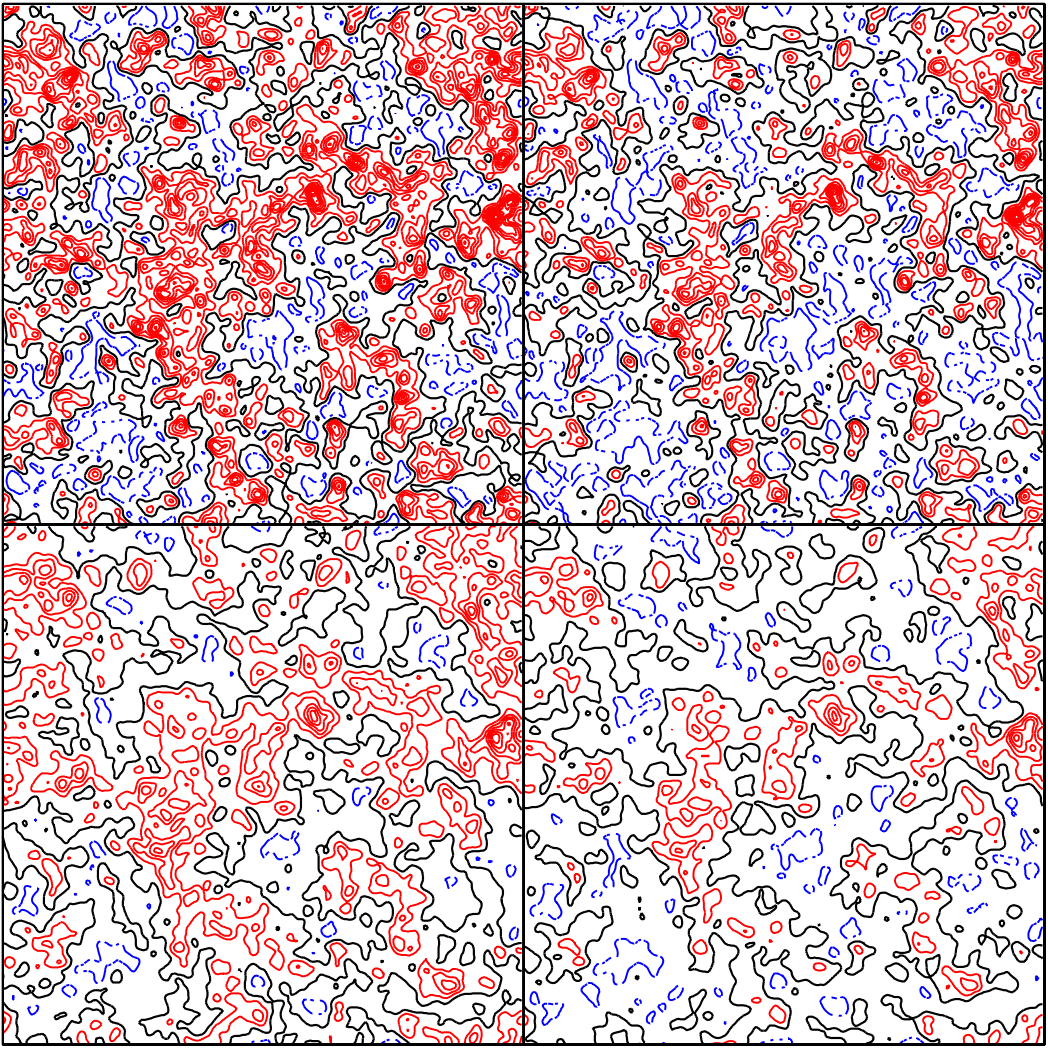,height=6.5in,width=6.5in}}
\caption{The same as the previous figure but for the gas density field with
$\sigma_{\delta}=0.6$. The contour spacing in this map is $3\rm \; mK$. }
\label{fig:tmap06}
\end{figure}

Figs \ref{fig:tmap1} and \ref{fig:tmap06} are contour maps 
of the DBT in the plane of the sky for gas rms density fluctuations of  $\sigma_{\delta}=1$ and $\sigma_{\delta}=0.6$, respectively. All  maps in all panels of the two figures show the DBT 
in the same slice of the box for a beamsize of $\Delta \theta=3.4$ arcmin.
In each figure, the panels in the top row  are for  $\Delta \nu=10\rm \; KHz$, while 
the bottom row is  for
$\Delta_\nu=100\rm \; KHz$.
The panels in the column to the left are  
 for neutral fraction (at mean density) 
of $\xhi=0.8$, while the column to the right shows maps for  $\xhi=0.9$.
As a quantitative guide to the map we list, in table (\ref{tab:tb}), 
 the mean,  $<\delta \tb>$, and  the $rms$, $< \delta \tb^{2} >^{{1/2}}$. 
A visual comparison between 
 the top ($\Delta \nu=10\rm \; KHz$) and
bottom ($\Delta \nu=100\rm \; KHz$) panels in the  figures   (\ref{fig:tmap1}) and (\ref{fig:tmap06}) reveal substantial differences. 
The fluctuations in the top panels are significantly higher and 
appear to show more structure.  
The  comoving physical scale of a $\Delta \theta=3.4$ arcmin is about 6.5 $\hmpc$, nearly 3 times larger than the physical scale corresponding to  
$\Delta \nu=100$ KHz.
Despite the large beamsize,  small scale fluctuations are clearly seen 
for $\Delta \nu=10$ KHz.

There are clear differences also between 
the contour maps for $\xhi=0.9$  (panels to the 
right) and for $\xhi=0.8$. These are caused by the electron-atom collisions
which are more effective for $\xhi=0.8$. The higher  fluctuations amplitude
seen in the panels to the left  is consistent with figures (\ref{fig:one0})
and (\ref{fig:one}).

The amplitude of the gas density fluctuations clearly plays an important
role in determining the level of DBT fluctuations. 
According to table (\ref{tab:tb}) the DBT fluctuations for  $\sigma_{\delta}=1$ 
are about 50\% higher than in the $\sigma_{\delta}=0.6$ case (see also figures (\ref{fig:tmap1}) and (\ref{fig:tmap06})).  
However, 
the mean values of  $\delta \tb$
are smaller in the former case.   The reason is the larger volume of 
space  occupied by under-dense regions  for  $\sigma_{\delta}=1$.

\begin{figure}
\centering
\mbox{\psfig{figure=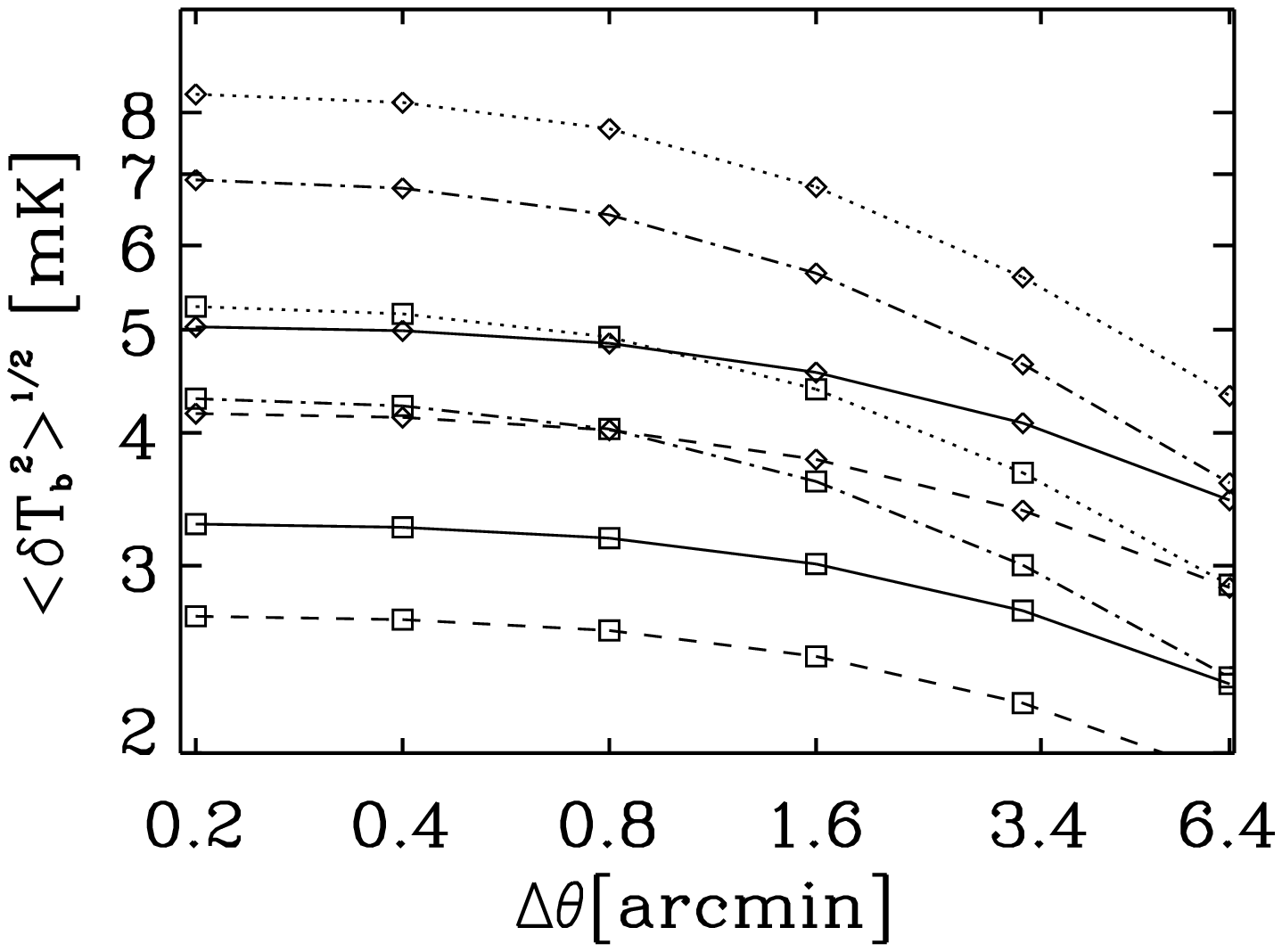,height=3.5in,width=5in}}
\caption{The $rms$ value of the DBT. The circles are attached to curves 
obtained from the density field with $\sigma_{\delta}=1$, while the squares 
are for the fields with $\sigma_{\delta}=0.6$. Solid, dotted, dashed and dash-dotted lines, respectively,  correspond
to $(\Delta \nu, \xhi)=(100{\rm \; KHz}, 0.8)$, 
(10,0.8), (100,0.9), and (10,0.9).
}
\label{fig:rms_theta}
\end{figure}

In figure (\ref{fig:rms_theta}) we plot the $rms$ of the DBT, $< \delta \tb^{2} >^{{1/2}}$, as a function of the beamsize, $\Delta \theta$.  The curves correspond to various cases as
explained in the figure caption. The $rms$ values as a function of  $\Delta \theta$ 
are  closely related to the average of angular correlation function within an 
angular separation of $\sim \Delta \theta$ (e.g. Peebles 1980).  None of the curves reveal any  characteristic feature related to a particular physical
scale.  
 This is not surprising since no such
 scale can be identified in
the X-ray pre-reionization scenario employed here. With the exception of the cases
corresponding to  $\Delta \nu =100\rm \; KHz$ with $\sigma_{\delta}=0.6$ (the solid
and dashed lines  with the square symbols), all  curves show a non-negligible
level of fluctuations. However, an inspection of the contour maps 
in figures (\ref{fig:tmap1}) and (\ref{fig:tmap06}) shows that the  DBT
fluctuations between  some regions can significantly be larger that the $rms$ 
values.  

\begin{table}
\centering
\caption{The mean and $rms$ value of the DBT for $\xhi=0.8$ and 0.9, as computed from the gas density fields with $\sigma_{\delta}=1$ and $\sigma_{\delta}=0.6$. In each entry the first number in the bracket is the mean $<\delta \tb>$ and the second is the $rms$ $< \delta \tb^{2} >^{{1/2}}$.}
\begin{tabular}{l|c|c|} \hline
 & $\xhi=0.8$    & $\xhi=0.9$    \\ \hline
   $\sigma_{\delta}=1$ & (9.05, 8.94)    & (7.86, 7.44)    \\ \hline
   $\sigma_{\delta}=0.6$ & (9.29,5.60)    & (8.05, 4.59)    \\ \hline
\end{tabular}
\label{tab:tb}
\end{table}

\begin{figure}
\centering
\mbox{\psfig{figure=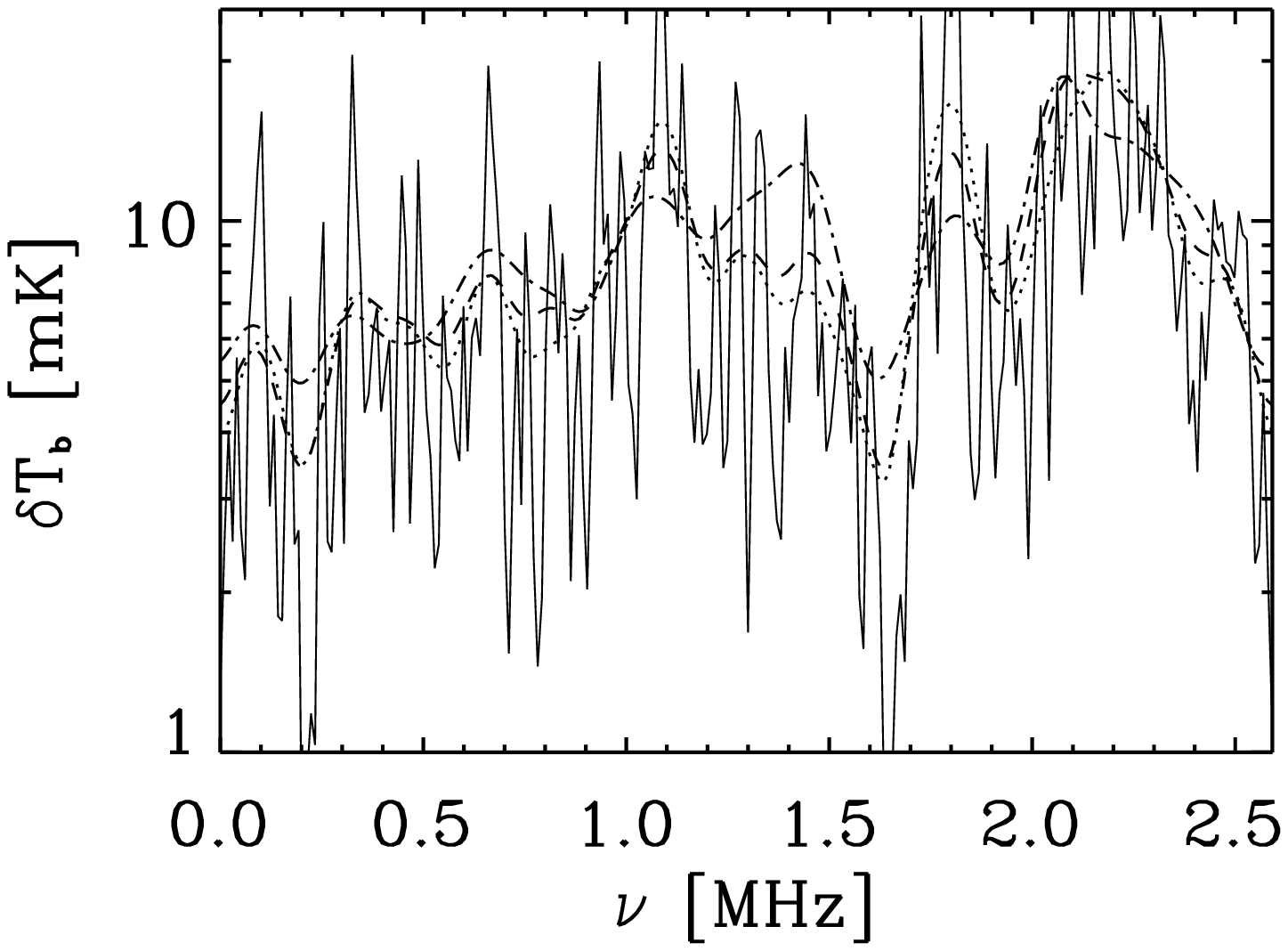,height=3.5in,width=5in}}
\mbox{\psfig{figure=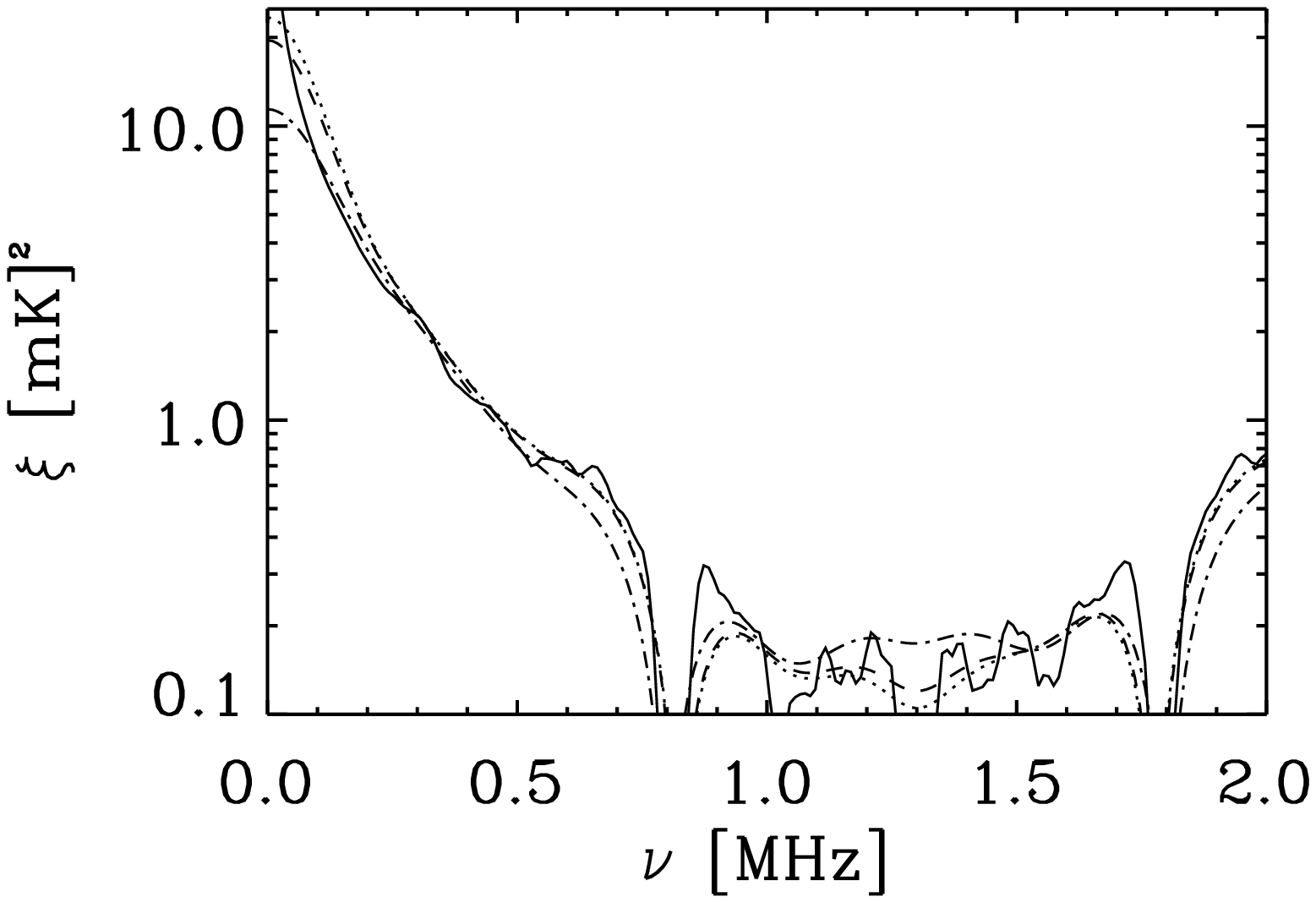,height=3.5in,width=5in}}
\caption{{\it Top}:  The DBT fluctuations in  a segment equivalent to 
 $L=40\hmpc$ in the frequency axis.
All curves are for the same segment but as seen with different angular resolutions.
{\it Bottom}: Correlation functions computed from the DBT as a function of 
frequency. In both panels, all curves for frequency resolution of $\Delta \nu=100\rm \; KHz$, and the  solid, dotted, dashed, and dash-dotted
lines correspond to angular resolutions of $\Delta \theta=0,0.4,1.6$, and 6.4 arcmin, respectively. }
\label{fig:los}
\end{figure}

We turn now to the fluctuations along the frequency axis. 
These are of interest since  the correlation along the frequency axis
can be extremely useful for disentangling the cosmological signal from
foreground contamination (Gnedin \& Shaver 2004,  Zaldarriaga, Furlanetto \& Hernquist 2004). 
In the top panel of Fig.~\ref{fig:los} we plot the DBT fluctuations in a 
segment of $40 \hmpc$ through the numerical box. 
All curves show DBT in the same segment  for a bandwidth of $100\rm \; KHz$, but 
for different angular resolutions.
As is clear from the bottom panel, the DBT is significantly correlated 
up to separations corresponding to a few $\hmpc$.

\subsection{Brightness temperature in recombination bubbles}
In this paper we have focused on 21-cm emission induced by collisions  
during a pre-reionization period in which X-rays are the main 
source of ionizing photons. 
An alternative scenario would be pre-reionization by UV
radiation produced by short-lived sources (of lifetime
much shorter than the recombination time-scale).

In this scenario re-ionization begins with the production 
of individual ionized bubbles engulfing the sources.
While a bubble recombines electron-atom collisions decouple the 
spin and CMB temperature efficiently so that 
significant 21cm emission is possible.  
The sources also produce Ly$\alpha$ photons which 
can decouple the CMB from the spin temperature of \hi .
But, 
 if the filling factor of the bubbles is small and 
 the birth rate of these short lived sources is  low 
then   Ly$\alpha$ photons may not become 
important at the early stages of re-ionization.  
In a separate paper we will present a detailed analysis of a situation like where the
source are identified with massive first stars  
 (e.g. Omukai \& Nishi 1998, Abel, Bryan \& Norman 2000, Bromm, Coppi \& Larson 2002).
Here we only consider the differential antenna temperature 
in a region undergoing recombinations after it has been completely 
ionized by 
a short-lived source.

\begin{figure}
\centering
\mbox{\psfig{figure=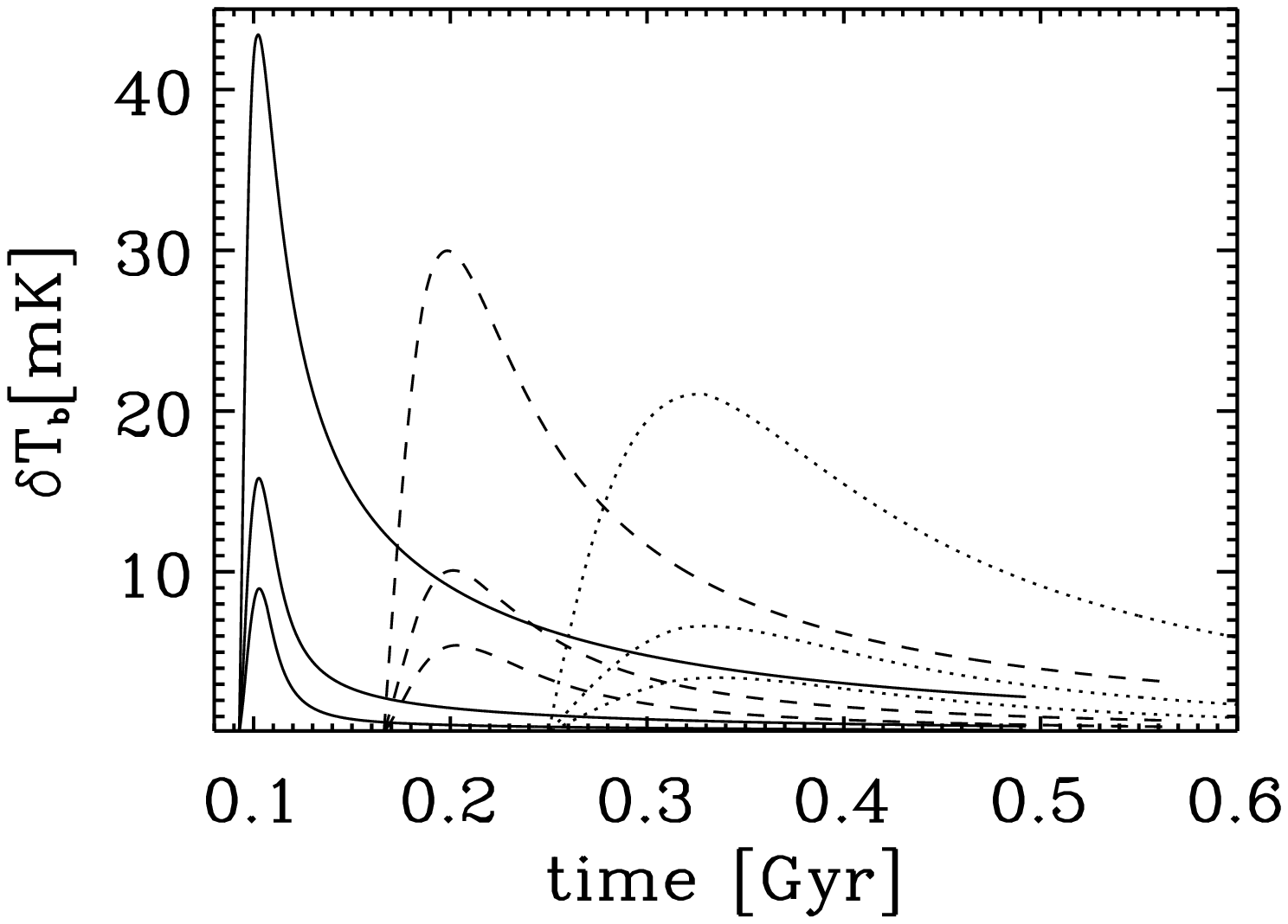,height=3.5in,width=5in}}
\caption{The DBT as a function of time in recombining regions assumed to 
have been fully ionized at redshift $z=30$ (solid curves), 20 (dashed), and
15 (dotted). 
The ionizing source is switched off during recombination. The three curves at each redshift, respectively,  correspond from top to bottom to mean density contrasts
of $\delta=1,$ 0, and $-0.3$ inside the bubbles.}
\label{fig:recomb}
\end{figure}

To obtain brightness temperature as a bubble recombines we have numerically integrated the 
recombination equations assuming  Compton cooling (e.g. Peebles 1993) to  
be the dominant cooler  at at the
relevant redshifts.
The recombination coefficients as a function of temperature are taken from Storey 
\& Hummer (1995). The de-excitation coefficients for atom-atom and atom-electron collisions are from Field (1958), which are in good agreement with other 
calculations in the literature for $T>100\rm \; K$ (Allison \& Dalgarno 1969).
In Fig.~\ref{fig:recomb} we plot $\delta \tb$ as a function 
of time for recombinations starting at $z=30$ (solid lines),
$z=20$ (dashed lines), and z=15 (dotted). The initial temperature after reionization 
is taken to be $T_{K}=10^{4}{\rm \; K}$. The three curves in each case correspond to different values of the density contrast inside the 
bubble: $\delta=1$ (higher  curve ), $\delta=0$ (middle curve), and $\delta=-0.3$ (lower curve). We have assumed uniform density inside the bubbles
in solving the  recombination
equations.
According to Fig.~\ref{fig:recomb} a recombining bubble can 
yield a significant 21-cm emission of $\delta \tb\sim 10 \rm \; mK$
at $\delta\gsim 0$ 
for about $10^{8}\rm Yr$ for bubbles generated at $z=15$ and 20.
At $z=30$ the emission lasts for a significant fraction of the Hubble
time at that redshift. Electron-atom collisions are clearly important
even for $\delta=-0.3$. Note that these bubbles can encompass large
volumes of space so that the mean density in them is likely to be very close to $\delta=0$. However, density variations inside a bubble
will yield  $\delta \tb$ fluctuations at the level of differences
between the curves shown in the figure.

\section{Discussion and conclusions} 

Recently,  the cosmological 21-cm signal from the epoch of reionization 
has been the subject of intensive  research activities. 
 Twenty one cm maps are  probably the only direct probe
of the structure of high redshift reionization. In comparison,  
secondary CMB anisotropies will provide constraints on line of sight integrated
quantities related to the ionized rather than neutral gas.
Also, the statistical properties of 21-cm fluctuations contain 
 valuable information on the physical nature of the ionizing sources and  
 the  
amplitude of the underlying mass at high redshift. 
Further, it has become clear that the prospects for measuring the 21-cm signal are good in view of 
the various radio telescopes designed with this purpose in mind. 
Despite foreground contaminations (e.g. Oh \& Mack 
2003, Di Matteo, Ciardi \& Miniati 2004, Morales \& Hewitt 2004, Zaldarriaga, Furlanetto \& Hernquist 2004), the collective data obtained by telescopes 
like the Low Frequency Array \footnote{http://www.lofar.org} (LOFAR), 
the Primeval Structure Telescope \footnote{http://astrophysics.phys.cmu.edu/~jbp}
(PAST), 
the Square Kilometer Array\footnote{http://www.skatelescope.org}, and 
the T-Rex instrument\footnote{http://orion.physics.utoronto.ca/sasa/Download/Papers/poster/poster.jpg}
should, at the very least, help us discriminate among  distinct reionization
scenarios.

In some reionization scenarios,
collisions of free electrons with \hi\ atoms  
can lead to a significant 21-cm cosmological 
signal from regions with density contrasts $(1+\delta)(\xhi+ {\tilde y}_{_{\rm e}}/{\tilde y}_{_{\rm HI}}(1-\xhi))>20[(1+z)/10]^{-2} $ and $\xhi\gsim 0.8$, where 
the  ${\tilde y}_{_{\rm e}}/{\tilde y}_{_{\rm HI}}$ is in the range 
$10-20$ (Field 1958).
We chose a simplified version of the  X-ray pre-reionization scenario  as an example for computing the fluctuations in the 21-cm brightness temperature. 
In this scenario  decoupling by collisions is likely 
to play a major role, at least sufficiently far from the sources.
We also demonstrated that collisions  are also important
during the recombination of gas in bubbles ionized by short
lived sources. 
A recombining bubble should 
show in 21-cm emission (of $\delta \tb\sim 10 \rm \; mK$)
for about $10^{8}\rm Yr$.
 A thorough analysis of the collective 
emission from these bubbles 
is currently underway.

In order to compute the fluctuations in the X-ray scenario, we have employed 
a simple semi-analytic model that allows us to assess
the statistical properties of 21-cm brightness temperature fluctuations.
The 21-cm fluctuations in this and other pre-reionization scenarios 
have been considered previously  by several workers in the field (Bruscoli \etal 2000, Benson \etal 2001, Liu \etal 2001, Iliev \etal 2002, 
Ciardi \& Madau 2003, Bharadwaj \& Sk. Saiyad 2004, Ricotti, Ostriker \& Gnedin 2004,
Zaldarriaga, Furlanetto \& Hernquist 2004). 
In particular Ricotti, Ostriker \& Gnedin (2004) used numerical simulations
of a $1\hmpc^{3}$ box to model various details of X-ray ionization
scenarios. However, an assessment of the importance of collisions vs Ly$\alpha$ pumping
as a function  of position in the simulation remains to be done.
Further, the simulation box is still small and an analysis of 
the general statistical properties of the fluctuations must rely on 
semi-analytic models.

We advocate here the 
the application of the Alcock-Paczy\'nski (AP) test  
(Alcock  \& Paczy\'nski  1979) on the correlations
measured in three dimensional maps of 21-cm emission. 
These correlations are readily expressed in terms of angular separations and 
frequency intervals. In order to derive the correlations in terms of real space separations (measured for example in $\hmpc$) one needs to know 
the mean density parameter, $\Omega_{\rm m}$, and the cosmological 
constant $\Omega_{\lambda}$. Constraints on these  parameters can then be  
obtained by demanding isotropy of correlations in real space.
A detailed analysis of the 21 cm AP test is presented elsewhere (Nusser 2004).


\section{Acknowledgment}

The author has benefitted from stimulating discussions with  J.P. Ostriker, M.J. Rees, P. Storey, and 
S. Zaroubi. He wishes to  thank the Department of
Physics and Astronomy, University College London, for the hospitality and support.
This work is supported by the Research and Training Network ``The physics of the Intergalactic Medium'' set up by the European Community under the contract HPRNCT-2000-00126 and by the German Israeli Foundation for the Development of 
Research.



\end{document}